\documentclass[12pt,a4paper]{article}
\usepackage{amssymb}
\usepackage{graphicx,color}
\makeatletter
\def\rddots{\mathinner{\mkern1mu\raise\p@%
    \vbox{\kern7\p@\hbox{.}}\mkern2mu%
    \raise4\p@\hbox{.}\mkern2mu\raise7\p@\hbox{.}\mkern1mu}}
\makeatother
\makeatletter
\def\eqnarray{%
\stepcounter{equation}%
\let\@currentlabel=\theequation
\global\@eqnswtrue
\global\@eqcnt\z@
\tabskip\@centering
\let\\=\@eqncr
$$\halign to \displaywidth\bgroup\@eqnsel\hskip\@centering
$\displaystyle\tabskip\z@{##}$&\global\@eqcnt\@ne
\hfil$\displaystyle{{}##{}}$\hfil
&\global\@eqcnt\tw@$\displaystyle\tabskip\z@{##}$\hfil
\tabskip\@centering&\llap{##}\tabskip\z@\cr}
\makeatother
\newcommand{\ket}[1]{{\vert{#1}\rangle}}

\begin{document}

\title{\sl Quantum Mechanics with Complex Time : \\
A Comment to the Paper by Rajeev}
\author{
  Kazuyuki FUJII
  \thanks{E-mail address : fujii@yokohama-cu.ac.jp }\\
  Department of Mathematical Sciences\\
  Yokohama City University\\
  Yokohama, 236--0027\\
  Japan
  }
\date{}
\maketitle
%
%
%
%
\begin{abstract}
  In (quant--ph/0701141) Rajeev studied quantization of the damped simple 
  harmonic oscillator and introduced a complex--valued Hamiltonian (which 
  is normal). In this note we point out that the quantization is 
  interpreted as a quantum mechanics with {\bf complex time}. We also 
  present a problem on quantization of classical control systems.
\end{abstract}
%


%
%
%
%

In the paper \cite{SGR} Rajeev studied a quantization of the damped 
simple harmonic oscillator and introduced a complex--valued Hamiltonian 
(which is normal). In the case where dissipative systems are treated 
complex--valued Hamiltonians cannot be avoided in general.

In this note we point out that the quantization with complex--valued 
Hamiltonian is interpreted as a quantum mechanics with complex time, 
which is in a certain sense reasonable. 

Anyway, quantization of dissipative systems must be widely studied. 

\par \vspace{5mm} \noindent 
\begin{Large}
{\bf Quantum Mechanics with Complex Hamiltonian}
\end{Large}

\par \vspace{5mm} \noindent 
First of all let us review the result in \cite{SGR} within our necessity. 
In \cite{SGR} there are some typos, so we give an explicit expression 
as much as we can. 

The (differential) equation of the damped simple harmonic oscillator is given 
by
\begin{equation}
\label{eq:starting equation}
\ddot{x}+2\gamma \dot{x}+\omega^{2}x=0,\quad \gamma > 0
\end{equation}
where $x=x(t),\ \dot{x}=dx/dt$ and the mass is set to 1 for simplicity. 
In the following we treat only the case $\omega > \gamma$ (the case 
$\omega=\gamma$ may be interesting).

\par \noindent
The general solution is well--known to be 
\begin{equation}
\label{eq:general solution}
x(t)=A\mbox{e}^{-\gamma t}\sin(\omega_{1}t+\theta)
\end{equation}
where $A$ and $\theta$ are real constants (see (\ref{eq:lambdas}) as to 
$\omega_{1}$).

A comment is in order.\ \ If $\omega=\gamma$ or $\omega_{1}=0$ then 
the solution is 
\[
x(t)=\mbox{e}^{-\gamma t}(c_{0}+c_{1}t)
\]
where $c_{0}$ and $c_{1}$ are real constants. However, in the paper we don't 
treat this case.

Let us rewrite the equation (\ref{eq:starting equation})
\begin{equation}
\dot{x}=p,\quad \dot{p}=-\omega^{2}x-2\gamma p
\end{equation}
or in the matrix form 
\begin{equation}
\label{eq:matrix form}
\frac{d}{dt}
\left(
  \begin{array}{c}
    x \\
    p
  \end{array}
\right)=
\left(
  \begin{array}{cc}
    0 & 1 \\
    -\omega^{2} & -2\gamma
  \end{array}
\right)
\left(
  \begin{array}{c}
    x \\
    p
  \end{array}
\right)\ \Longleftrightarrow\ \frac{d}{dt}X=AX.
\end{equation}
Next, let us make $A$ diagonal. The characteristic equation of $A$ is
\[
0=|\lambda E-A|
=
\left|
  \begin{array}{cc}
    \lambda & -1 \\
    \omega^{2} & \lambda+2\gamma
  \end{array}
\right|
=(\lambda +\gamma)^{2}+\omega^{2}-\gamma^{2}
\equiv
(\lambda +\gamma)^{2}+{\omega_{1}}^{2}
\]
, from which we have 
\begin{equation}
\label{eq:lambdas}
\lambda_{-}=-\gamma-i\omega_{1},\quad 
\lambda_{+}=-\gamma+i\omega_{1};
\qquad \omega_{1}=\sqrt{\omega^{2}-\gamma^{2}}.
\end{equation}
By the usual procedure, if we set
\[
U=\frac{1}{\sqrt{2\omega_{1}}}(|\lambda_{-}),|\lambda_{-}))
=\frac{1}{\sqrt{2\omega_{1}}}
\left(
  \begin{array}{cc}
    1           & 1           \\
    \lambda_{-} & \lambda_{+}
  \end{array}
\right)
\ \Longrightarrow\ 
U^{-1}
=\frac{-i}{\sqrt{2\omega_{1}}}
\left(
  \begin{array}{cc}
    \lambda_{+}  & -1 \\
    -\lambda_{-} & 1
  \end{array}
\right)
\]
then
\[
A=U
\left(
  \begin{array}{cc}
    \lambda_{-} &             \\
                & \lambda_{+}
  \end{array}
\right)
U^{-1}\equiv UD_{A}U^{-1}
\]
and from (\ref{eq:matrix form})
\[
\frac{d}{dt}X=AX\ \Longrightarrow\ 
\frac{d}{dt}U^{-1}X=D_{A}U^{-1}X.
\]
Therefore new variables are 
\begin{equation}
\label{eq:new variables}
\left(
  \begin{array}{c}
    z     \\
    z^{*}
  \end{array}
\right)
\equiv
U^{-1}
\left(
  \begin{array}{c}
    x \\
    p
  \end{array}
\right)
=
\left(
  \begin{array}{c}
    \frac{1}{\sqrt{2\omega_{1}}}(\omega_{1}x+i(p+\gamma x)) \\
    \frac{1}{\sqrt{2\omega_{1}}}(\omega_{1}x-i(p+\gamma x))
  \end{array}
\right)
\end{equation}
and
\begin{equation}
\label{eq:new equations}
\frac{dz}{dt}=\lambda_{-}z,\quad \frac{dz^{*}}{dt}=\lambda_{+}z^{*}.
\end{equation}
The equations have some deep meaning as shown in the latter.

The Poisson bracket is defined as
\begin{equation}
\{F,G\}=\frac{\partial F}{\partial x}\frac{\partial G}{\partial p}-
\frac{\partial G}{\partial x}\frac{\partial F}{\partial p}
\end{equation}
for $F=F(x,p)$ and $G=G(x,p)$. This gives $\{x,p\}=1$, and therefore 
$\{z^{*},z\}=i\{x,p\}=i$ from (\ref{eq:new variables}). 
Here, if we define a complex--valued Hamiltonian
\begin{equation}
\label{eq:complex--valued Hamiltonian}
{\cal H}=(\omega_{1}-i\gamma)zz^{*}
\end{equation}
then the equations (\ref{eq:new equations}) are easily recovered
\[
\frac{dz}{dt}=\lambda_{-}z=-i(\omega_{1}-i\gamma)z=\{z,{\cal H}\},
\quad 
\frac{dz^{*}}{dt}=\lambda_{+}z^{*}=i(\omega_{1}+i\gamma)z^{*}=
\{z^{*},{\cal H}^{*}\}.
\]
In this stage we cannot avoid a complex Hamiltonian.

A comment is in order.\ \ The limit $\gamma \rightarrow 0$ recovers 
the usual Hamiltonian ${\cal H}=\omega zz^{*}$, while the limit 
$\gamma \rightarrow \omega$ gives ${\cal H}=-i\omega zz^{*}$ (pure 
imaginary).

Next, let us turn to the canonical quantization of the system. The usual 
procedure is given by the correspondence (representation) 
$z \rightarrow a^{\dagger}$ and $z^{*} \rightarrow \hbar a$ 
with $[a,a^{\dagger}]=1$. Then the (quantized) Hamiltonian becomes 
\begin{equation}
\label{eq:quantized Hamiltonian}
{\cal H}=(\omega_{1}-i\gamma)zz^{*}=(\omega_{1}-i\gamma)
\frac{zz^{*}+z^{*}z}{2}\longrightarrow 
H=\hbar(\omega_{1}-i\gamma)(a^{\dagger}a+1/2),
\end{equation}
from which the energy spectrum appears to be
\[
\hbar(\omega_{1}-i\gamma)(n+1/2)=
\hbar(\omega_{1}-i\gamma)n+\frac{\hbar(\omega_{1}-i\gamma)}{2}
\]
for $n \geq 0$. However, this is not true. We must impose a restriction 
on the ground state energy. 
Namely, it is free of $\gamma$, which is reasonable from the physical point 
of view. Therefore 
\[
{
\frac{\hbar(\omega_{1}-i\gamma)}{2}
\left|
  \begin{array}{c}
    {}\\
  \end{array}
\right. \hspace{-3mm}
}_{\gamma=0}=\frac{\hbar \omega}{2},
\]
so the real spectrum\footnote{This point is not clear in \cite{SGR}} is
\begin{equation}
\label{eq:real spectrum}
\hbar(\omega_{1}-i\gamma)n+\frac{\hbar \omega}{2}
\quad \mbox{for}\quad n \geq 0.
\end{equation}

Let us consider the time evolution of our system. For any initial state
\[
\ket{\psi}=\sum_{n=0}^{\infty}\psi_{n}\ket{n}\quad \mbox{where}\quad 
\sum_{n=0}^{\infty}|\psi_{n}|^{2}=1
\]
the time evolution is given by
\begin{equation}
\label{eq:time evolution}
\ket{\psi}\longrightarrow \ket{\psi(t)}\equiv \mbox{e}^{-itH}\ket{\psi}
=\mbox{e}^{-it\frac{\hbar \omega}{2}}
\sum_{n=0}^{\infty}\psi_{n}\mbox{e}^{-it\hbar(\omega_{1}-i\gamma)n}\ket{n}
\end{equation}
from (\ref{eq:real spectrum}). Since
\[
\sum_{n=0}^{\infty}\psi_{n}\mbox{e}^{-it\hbar(\omega_{1}-i\gamma)n}\ket{n}
=
\sum_{n=0}^{\infty}
\psi_{n}\mbox{e}^{-t\hbar\gamma n}\mbox{e}^{-it\hbar\omega_{1}n}\ket{n},
\]
the limit $t\rightarrow \infty$ gives
\begin{equation}
\ket{\psi(t)}\longrightarrow 
\mbox{e}^{-it\frac{\hbar \omega}{2}}\psi_{0}\ket{0}.
\end{equation}
All states will settle down to the ground state after infinite time. 
Compared with the classical result (\ref{eq:general solution}) 
all this sounds reasonable as Rajeev says.

\par \vspace{10mm} \noindent 
\begin{Large}
{\bf Quantum Mechanics with Complex Time}
\end{Large}

\par \vspace{5mm} \noindent 
In this section we point out that it is possible to transform the preceding 
result into the usual form with complex time.

From (\ref{eq:new equations}) with (\ref{eq:lambdas}) we have
\[
\frac{dz}{dt}=(-\gamma-i\omega_{1})z=
\left(1-i\frac{\gamma}{\omega_{1}}\right)(-i\omega_{1})z,\ \ 
\frac{dz^{*}}{dt}=(-\gamma+i\omega_{1})z^{*}=
\left(1+i\frac{\gamma}{\omega_{1}}\right)(i\omega_{1})z^{*}.
\]
From this it is reasonable to introduce a complex time
\begin{equation}
\label{eq:complex time}
\tau=\left(1-i\frac{\gamma}{\omega_{1}}\right)t
\ \Longrightarrow\ 
\tau^{*}=\left(1+i\frac{\gamma}{\omega_{1}}\right)t,
\end{equation}

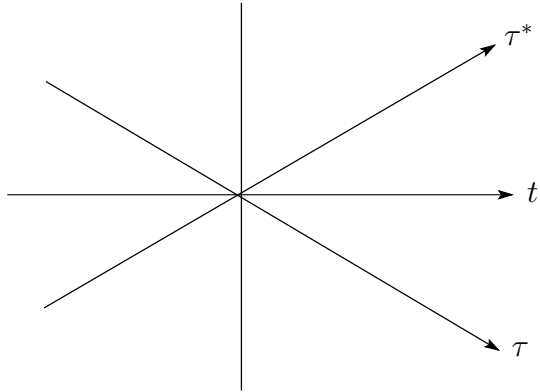
\vspace{5mm}
\begin{center}
\input{complex-time.fig}
\end{center}

\par \noindent
so the equations above reduce to
\begin{equation}
\frac{dz}{d\tau}=-i\omega_{1}z,\quad
\frac{dz^{*}}{d\tau^{*}}=i\omega_{1}z^{*}.
\end{equation}

If we define a (usual) Hamiltonian
\begin{equation}
\widetilde{\cal H}=\omega_{1}zz^{*}
\end{equation}
then we can recover
\[
\frac{dz}{d\tau}=-i\omega_{1}z=\{z,\widetilde{\cal H}\},\quad
\frac{dz^{*}}{d\tau^{*}}=i\omega_{1}z^{*}=\{z^{*},\widetilde{\cal H}\}.
\]
Therefore the usual system of harmonic oscillator is obtained 
except for complex time. 

The quantum Hamiltonian is 
\begin{equation}
\widetilde{H}=\hbar\omega_{1}(a^{\dagger}a+1/2)
\end{equation}
and the ``time" evolution is
\begin{eqnarray}
\label{eq:new time evolution}
&&\ket{\psi}\longrightarrow 
\ket{\psi(\tau)}\equiv \mbox{e}^{-i\tau \widetilde{H}}\ket{\psi}
=\mbox{e}^{-it\frac{\hbar \omega}{2}}
\sum_{n=0}^{\infty}\psi_{n}\mbox{e}^{-i\tau\hbar\omega_{1}n}\ket{n}, \\
&&\ket{\psi}\longrightarrow 
\ket{\psi(\tau^{*})}\equiv \mbox{e}^{i\tau^{*} \widetilde{H}}\ket{\psi}
=\mbox{e}^{it\frac{\hbar \omega}{2}}
\sum_{n=0}^{\infty}\psi_{n}\mbox{e}^{i\tau^{*}\hbar\omega_{1}n}\ket{n}
\end{eqnarray}
for $\ket{\psi}=\sum_{n=0}^{\infty}\psi_{n}\ket{n}$, where we again assumed
that the ground state energy is independent of $\gamma$ (or independent of 
imaginary part, Re $\tau=\mbox{Re}\ \tau^{*}=t$).

Again, all this sounds reasonable.

\vspace{5mm}
Now, we present a problem. One of the simplest examples of classical 
control system is
\begin{equation}
\ddot{x}+2\gamma \dot{x}+\omega^{2}x=f(t),\quad \gamma > 0
\end{equation}
where $f(t)$ is some outer field that we can control. In the matrix form
\begin{equation}
\frac{d}{dt}
\left(
  \begin{array}{c}
    x \\
    p
  \end{array}
\right)=
\left(
  \begin{array}{cc}
    0 & 1 \\
    -\omega^{2} & -2\gamma
  \end{array}
\right)
\left(
  \begin{array}{c}
    x \\
    p
  \end{array}
\right)+
f(t)
\left(
  \begin{array}{c}
    0 \\
    1
  \end{array}
\right).
\end{equation}

Since the system has been quantized if $f(t)=0$, construct a theory 
to control the quantized system under $f(t)$. Note that $f(t)$ is a classical 
field and not a quantized one.

\vspace{5mm}
In last, let us conclude the paper by stating the motivation. 
We are studying a quantum computation based on Cavity QED, see for example 
\cite{FHKW1} and \cite{FHKW2}. To construct a more realistic model of 
quantum computer we must take {\bf decoherence} into consideration, which is 
however a very hard problem. How to control decoherence has not been known 
as far as we know. Much work is required !

\vspace{10mm}
\noindent{\em Acknowledgment.}\\
The author wishes to thank K. Funahashi for helpful comments and suggestions.

\vspace{5mm}

\end{document}